\documentclass[aps,pra,reprint,superscriptaddress]{revtex4-2}
\usepackage{graphicx}
\usepackage{dcolumn}
\usepackage{bm}
\usepackage{subfig}  
\usepackage{caption}  
\usepackage{ragged2e} 
\usepackage{hyperref}
\usepackage{amsmath}
\usepackage{esint}
\usepackage{float}
\usepackage{orcidlink}
\usepackage{lineno}

\begin{document}
	
\preprint{APS/123-QED}
	
\title{Experimental Determination of Slow-Neutron Detection Efficiency and Background Discrimination in Mixed Radiation Fields Using Differential CR-39 Track Detectors}
	
\author{Ankit Kumar\,\orcidlink{0000-0001-5962-7914}}
\email{ankitkumar5111994@gmail.com, ankt@iitk.ac.in}
\affiliation{Department of Physics, Indian Institute of Technology Kanpur, Kanpur, India}
	
\author{Tushar Verma\,\orcidlink{0009-0005-1810-6561}}
\affiliation{Department of Chemical Engineering, Indian Institute of Technology Kanpur, Kanpur, India}
	
\author{Pankaj Jain\,\orcidlink{0000-0001-8181-5639}}
\affiliation{Space, Planetary \& Astronomical Sciences \& Engineering, Indian Institute of Technology Kanpur, Kanpur, India}
	
\author{Raj Ganesh Pala\,\orcidlink{0000-0001-5243-487X}}
\affiliation{Department of Chemical Engineering, Indian Institute of Technology Kanpur, Kanpur, India}
	
\author{K.~P.~Rajeev\,\orcidlink{0000-0002-4685-6766}}
\email{kpraj@iitk.ac.in}
\affiliation{Department of Physics, Indian Institute of Technology Kanpur, Kanpur, India}

\begin{abstract}
	Accurate quantification of slow-neutron flux in mixed radiation environments containing charged-particle and fast-neutron backgrounds is critically important, particularly in systems where the radiation-field composition is not known \textit{a priori}. In this work, we present an experimentally validated method for the quantitative determination of slow-neutron fluence using CR-39 (Columbia Resin-39) solid-state nuclear track detectors. The technique is based on a differential detection configuration employing paired boron-coated CR-39 (BCR) and uncoated CR-39 (CCR) detectors, exploiting the large thermal-neutron capture cross section of boron together with the intrinsic insensitivity of bare CR-39 to slow neutrons.
	
	The CCR detector records tracks generated by charged particles and fast-neutron interactions, whereas the BCR detector additionally registers charged particles emitted following the
	$^{10}\mathrm{B}(n,\alpha)^{7}\mathrm{Li}$ reaction. Consequently, the difference in areal track densities between the paired detectors provides a conservative and physically well-defined measure of slow-neutron fluence.
	
	Controlled irradiation experiments demonstrate that the BCR--CCR differential signal increases linearly with exposure time. Analysis based on Poisson counting statistics yields a fitted slow-neutron--equivalent track rate of
	$R = 5.84 \pm 0.18~\text{tracks\,min}^{-1}$,
	which is significantly different from zero at the $32\sigma$ confidence level. The extracted response exhibits no evidence of curvature, temporal dependence, or saturation over the investigated exposure range, confirming that the BCR--CCR configuration provides a stable and reproducible basis for quantitative slow-neutron measurements. The detection efficiency of the BCR detector is further evaluated from the measured slope. The CCR track density measured in this study provides a quantitative proxy for the fast-neutron component leaking from the thermal-neutron source employed in the experiment.
	
	These results demonstrate that boron-integrated CR-39 detectors used in isolation are insufficient for accurate quantitative determination of slow-neutron fluence in mixed radiation fields. Reliable measurement requires the simultaneous deployment of an uncoated CR-39 control detector, with the BCR--CCR differential signal constituting the appropriate and conservative estimator of absolute thermal-neutron flux. Within this framework, the CCR response represents the non-slow-neutron background contribution. The method is particularly effective in experimental environments where the radiation-field composition is unknown and where the use of conventional active neutron detectors is impractical.

\end{abstract}

\maketitle
\section{Introduction}

Reliable measurement of thermal neutron fluences is of fundamental importance in nuclear physics, radiation dosimetry, materials characterization, plasma diagnostics, and laboratory-based neutron studies. In many experimental environments, neutron fluxes are weak, spatially nonuniform, or embedded within mixed-energy or mixed-radiation fields, rendering their accurate quantification experimentally challenging. Conventional active neutron detectors, such as $^3$He proportional counters, BF$_3$ tubes, and scintillator-based systems, being active detectors, are costly, in limited supply~\cite{KOUZES20101035,refId0}, and often face practical limitations due to inherent electronic noise when used for detecting neutron fluxes comparable to background levels~\cite{Langford_2013,rozov2010monitoringthermalneutronflux}. Other limitations include restricted geometrical flexibility, additional the requirement of neutron moderators, and incompatibility with operation inside complex media such as liquids and plasma systems.
\\

CR-39 solid-state nuclear track detectors~\cite{durrani2013solid} provides a cost-effective passive detector for nuclear diagnostics under such conditions and represent one of the most extensively studied polymer detectors, having been widely employed for neutron detection~\cite{nassiri2025optimization,TSURUTA01111992}, charged-particle detection~\cite{he2020calibration}, and heavy-ion detection~\cite{schollmeier2023differentiating}. When traversed by energetic charged particles with linear energy transfer (LET) exceeding its detection threshold $(\approx 15~\mathrm{keV}/\mu\mathrm{m})$~\cite{jadrnickova2006dosimetry}, CR-39 forms latent damage tracks with near-unit detection efficiency~\cite{seguin2003spectrometry}. These tracks can be chemically etched and are revealed aswell-defined microscopic pits, that are subsequently analyzed using optical microscopy. Owing to this threshold behavior, CR-39 is intrinsically insensitive to electromagnetic pulses~\cite{seguin2003spectrometry}, X rays and $\gamma$ rays ~\cite{rinderknecht2015impact}, $\beta$ particles, thermal neutrons, and other low-LET radiation~\cite{fleischer2022nuclear}. In addition, as a fully passive detector requiring no electronics, CR-39 is immune to electromagnetic interference and can be deployed in complex media with flexible detector geometries, including active volumes as small as $3\,\mathrm{mm} \times 3\,\mathrm{mm} \times 1.5\,\mathrm{mm}$.
\\

CR-39 based detectors has been utilized as a well-established neutron dosimeter over a broad energy range, from thermal energies $(\approx 25~\mathrm{meV})$ to fast neutrons of several tens of MeV $(\approx 66~\mathrm{MeV})$~\cite{izerrouken2003wide}. At fast-neutron energies, detection occurs predominantly via elastic scattering interactions with constituent hydrogen nuclei in the polymer matrix, producing recoil protons that form latent tracks in the detector material~\cite{durrani2013solid,izerrouken2003wide}. rIn contrast, slow or thermal neutrons detection requires the use of a neutron-to-charged-particle conversion layer. A widely adopted approach employs boron coating over CR-39, exploiting the large thermal-neutron capture cross section of the $^{10}\mathrm{B}(n,\alpha){}^{7}\mathrm{Li}$ reaction,in the thermal and epithermal energy range~\cite{durrani2013solid}. This reaction produces energetic $\alpha$ particles and ${}^{7}\mathrm{Li}$ nuclei that readily exceed the linear energy transfer (LET) threshold of the underlying CR-39 substrate and generate observable latent tracks on it. The areal particle track density observed after neutron exposure, thus provides quantitative information for fluence measurements.
\\

Such methods has been extensively utilized in diverse experimental contexts, including thermal-neutron dosimetry ~\cite{nassiri2025optimization}, investigations of boron neutron capture processes~\cite{smilgys2013boron}, and diagnostics of D–D fusion reactions ~\cite{PhysRevLett.112.095001}, detection of thermal neutrons generated during high-voltage atmospheric discharge phenomena ~\cite{agafonov2013observation}. 
\\

A key limitation in CR-39–based slow-neutron measurements is the possible simultaneous generation of charged particles and/or fast neutrons. These radiation components are also capable of producing tracks in boron-coated CR-39 detectors employed as slow-neutron sensors. As a result, tracks originating from non-thermal radiation may be incorrectly attributed to slow neutrons, leading to significant errors in the estimated slow-neutron fluence. This limitation becomes particularly critical in emerging experimental investigations of thermal neutrons in inertial confinement fusion systems~\cite{frenje2002absolute,frenje2020nuclear}, atmospheric discharge experiments~\cite{agafonov2013observation}, where mixed radiation fields are expected and reliable discrimination of neutron energy components remains unresolved. Consequently, this issue constitutes a significant research gap in the accurate characterization of thermal neutrons in such environments. Eliminating these ambiguities in experimental interpretation is therefore essential and can be achieved through the development of refined CR-39–based neutron measurement protocols.
\\

The present work addresses this practical research gap by establishing a consistent CR-39–based slow-neutron detection methodology capable of resolving slow neutron-induced signals at levels comparable to the intrinsic background, while simultaneously discriminating against contributions from the embedded charged particles and fast neutron backgrounds. To achieve this, a paired-detector approach is employed, consisting of a boric-acid-coated CR-39 (BCR) as a slow neutron sensor and an uncoated CR-39 control detector (CCR), deployed together at each detection site. Owing to its insensitivity to thermal neutrons, the CCR responds exclusively to charged particles and fast-neutron interactions, thereby providing an upper bound on such background contributions. Consequently, in this framework,  the difference in areal track densities measured on the BCR and CCR detectors yields a differential signal that constitutes a conservative estimate of the slow-neutron fluence as discussed later in this paper.
\\

In this paper, we first describe the preparation of paired BCR and CCR detectors, followed by a statistical evaluation of their response characteristics through controlled exposures to a calibrated plutonium–beryllium thermal neutron source. This analysis serves to validate the detection efficiency and reliability of the proposed methodology. All experimental raw data supporting the conclusions of this study are provided as Supplemental Material accompanying the main paper.
\\

The novelty of this study lies in establishing and experimentally validating a paired-detector CR-39 methodology that enables consistent and reliable discrimination between genuine slow-neutron signals and contributions arising from any other non-thermal radiation backgrounds, especially in experiments where the radiation field is not known a priori.

\section{Experimental Setup}


\subsection{Material and Equipments}

Columbia Resin-39 chips of size $6 $~mm$\times 4$~mm$ \times 1.5$~mm (Tastrak, Bristol, UK), $42~\mu$m thick polypropylene cello tape, $20~\mu$m thick polyethylene sheet, boric acid powder (Fischer Scientific), NaOH (Fischer Scientific), Deionized water and Leica DM 2700P microscope.

\subsection{Detector Preparation}

It is important to note that CR-39 detectors exhibit a finite track density even before irradiation. This intrinsic background, referred to as the baseline track density, corresponds to tracks present on the detector surface prior to any experimental exposure. In the present study, the baseline areal track density was measured to be $(11.6 \pm 0.7)$~tracks$\cdot$4.8~mm$^{-2}$ ~\cite{kumar2026cr39trackdetectorsignatures}. This value was subtracted from all CR-39–based measurements to avoid inclusion of pre-existing tracks in the analysis.

The CR-39–based boron-coated CR-39 (BCR) and control CR-39 (CCR) detectors were prepared using an identical procedure, differing only in the presence of the boron coating. For BCR preparation, boric acid was spread over a
$10~\mathrm{mm} \times 10~\mathrm{mm}$ adhesive region of a
$42~\mu\mathrm{m}$-thick polypropylene tape. After removal of excess material
by gentle air blowing, a visibly uniform boric acid layer remained adhered to the tape. To quantify the typical variation in the mass of boric acid coated on
polypropylene sheets ($10~\mathrm{mm} \times 10~\mathrm{mm}$ each), the masses
of ten tapes were measured before and after coating.The resulting coating masses were $1.3 \pm 0.1$, $1.3 \pm 0.1$, $1.4 \pm 0.1$, $1.2 \pm 0.1$, $1.3 \pm 0.1$, $1.4 \pm 0.1$, $1.4 \pm 0.1$, $1.3 \pm 0.1$, $1.3 \pm 0.1$, and $1.4 \pm 0.1~\mathrm{mg}$, with a mean of $1.3~\mathrm{mg}$ and a standard deviation of $0.1~\mathrm{mg}$,
indicating a small sample-to-sample variation. A $6~\mathrm{mm} \times 4~\mathrm{mm}$ piece was cut from such boric acid coated polypropylene sheet and affixed onto a bare CR-39 detector chip such that the boron layer is sandwiched between the polypropylene pouch and the CR-39 substrate. The resulting assembly was then sealed in a $20~\mu$m thick waterproof polyethylene pouch, resulting in BCR with a total effective plastic thickness of $65~\mu$m on the boron-coated/neutron detection surface. The CCR was prepared in the same manner but without boric acid coating, so the only difference between the BCR and CCR is the presence of the boron coating on the BCR. 
\\

Following the preparation of the boron-coated CR-39 (BCR) and control CR-39 (CCR) detectors, the track formation rates were independently evaluated at a location sufficiently distant from the thermal neutron source. The measured background values were $(3.39 \pm 0.10)$~tracks$\,(4.8~\text{mm}^2)^{-1}\,\text{week}^{-1}$
 for the BCR and $(1.25 \pm 0.11)$~tracks$\,(4.8~\text{mm}^2)^{-1}\,\text{week}^{-1}$
  for the CCR, respectively~\cite{kumar2026cr39trackdetectorsignatures}.It is noted that this contribution was found to be negligible over the maximum experimental duration of 40~minutes considered in the present study. Consequently, no explicit background correction for the BCR and CCR detectors is required, as will be demonstrated later. Furthermore, since the baseline track density is common to both BCR and CCR, its contribution cancels when the difference in track counts between the two detectors is evaluated. Therefore, no baseline correction is necessary for differential track analysis.

\subsection{Neutron Exposure}

The BCR and CCR detectors were exposed to calibrated thermal neutron source under fixed geometric alignment and shielding conditions. Exposure duration of 10, 20, 30, and 40 minutes were employed. All detector samples were positioned at identical locations relative to the source to maintain a uniform incident neutron flux across all measurements. The details of the reference thermal neutron source is provided in ~\cite{PURTY2017144}. 

\subsection{Chemical Etching and Track Counting}

Following irradiation, the CR-39 substrates of BCR and CCR detectors were obtained and chemically etched in preheated $6$~M NaOH souution in deionized H$_2$O, at $70 \pm 2~^\circ$C for 6~h After etching, samples were rinsed under tap water, gently blotted dry, and immediately imaged using an optical microscope across four areal regions, one at each corner, with each measuring $4.8$~mm$^2$ ($2.5$~mm$ \times 1.9$~mm), yielding four track measurements per CR-39 sample. Since all track images presented in this study are observed on a same area ($4.8$~mm$^2$) particle tracks per $4.8$~mm$^2$ are hereafter referred to simply as particle tracks. Any further details regarding experimental methods in provided in Sec~I.E.of the Supplememntal material of ~\cite{kumar2026cr39trackdetectorsignatures}.

\section{Results and Discussions}
Microscopic analysis of detectors exposed to a thermal neutron  reference field revealed significantly higher particle tracks on BCR compared to the adjacently placed CCR detector.

\begin{figure*}[!tb]
	\centering
	\includegraphics[width=\textwidth]{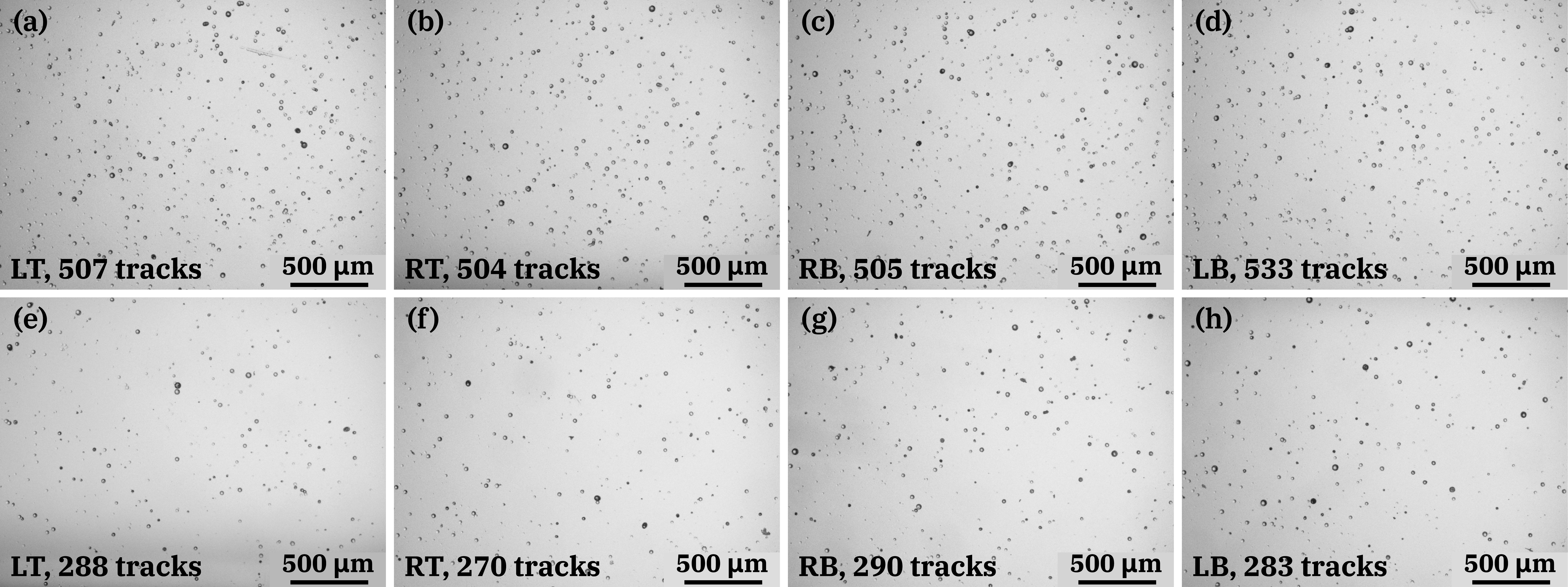}
	\caption{\justifying Optical micrographs of etched tracks on boron-coated CR-39 (BCR) and control CR-39 (CCR) detectors exposed to a thermal neutron source for 40 minutes. For each detector of 6~mm $\times$ 4~mm area, four equal non overlapping regions, each of area 1.8~mm $\times$ 2.5~mm (4.8~mm$^2$) were imaged: left-top (LT), right-top (RT), right-bottom (RB), and left-bottom (LB). Images (a)–(d) show BCR tracks counts of 507 (LT), 504 (RT), 505 (RB), and 533 (LB), yielding a mean of $512 \pm 6$ tracks per 4.8~mm$^2$. Images (e)–(h) show the corresponding CCR track counts of 288 (LT), 270 (RT), 290 (RB), and 283 (LB), yielding a mean of $283 \pm 4$ tracks per 4.8~mm$^2$. The actual number of mean tracks recorded during exposure on either detector is obtained after deduction of baseline value from each, which gives $500 \pm 6$ and $271 \pm 4$ for BCR and CCR respectively. Raw data is provided in Sec.~I of the Supplemental Material.}
	\label{fig:2}
\end{figure*}

For instance, during a 40 minute exposure, as shown in Figs.~\ref{fig:2}(a-d), four surveyed regions on BCR recorded an average of $500 \pm 6$ tracks while the adjacently placed CCR [see Figs.~\ref{fig:2}(e-h)] recorded an average of $271 \pm 4$ tracks, exhibiting measurable activity of both detcetors, with the BCR exhibiting an excess of $229 \pm 7$ tracks over CCR track counts (larger by a factor of $\sim 1.8$). 

In general, the CCR detector exhibits a higher detection efficiency for fast neutrons (and/or charged particles) than the BCR detector, since fast neutrons (and/or charged particles) may undergo scattering or partial attenuation in the boron conversion layer of the BCR, whereas no such attenuation occurs in the uncoated CCR. Consequently, for any exposure in which nuclear reactions between the incident radiation and the boron layer are absent, the CCR is expected to register a higher track density than the BCR. In contrast, an excess of tracks observed in the BCR relative to the CCR can arise only in the presence of a slow-neutron field, wherein neutron-capture reactions in the boron layer generate additional charged-particle tracks. The capture of slow neutrons by $^{10}$B produces $\alpha$ and $^{7}$Li particles via $^{10}$B(n, $\alpha$)$^{7}$Li, which form tracks in the underlying CR-39 of BCR and as the CCR contains no boron, no such response is possible, making the presence of slow neutron capture a unique explanation for the observed BCR excess of $229 \pm 7$ tracks over the CCR in 40 minute exposure experiment. 

We further note that even in the case where $\mathrm{BCR} > \mathrm{CCR}$ due to the presence of a slow-neutron field, the CCR may record/exhibit either zero or finite track densities, depending on whether fast neutrons (and/or charged particles) are embedded within the slow neutron irradiation field or not.

It is important to note that thermal neutron sources operating on the principle of moderating fast neutrons—such as those produced by natural fast-neutron emitters (e.g., plutonium--beryllium or californium-252)—reduce neutron energies primarily through elastic scattering with hydrogen nuclei in the moderator. Owing to incomplete moderation and neutron leakage processes, a residual fast-neutron component may remain present at the exposure location in addition to the thermal neutron flux.

Consequently, the finite number of tracks recorded on the CCR ($271 \pm 4$ tracks) represents an upper bound on the contribution from leaking fast neutrons at the detection site. More importantly, the differential track signal obtained from the paired detectors, $\mathrm{BCR}-\mathrm{CCR} = 229 \pm 7$ tracks, provides a conservative estimate of the slow-neutron fluence at the same location.
 
To further examine the properties of the BCR detectors exposed to slow neutron source, the detectors were irradiated for different exposure times (10, 20, and 30 minutes), in addition to the 40 minute exposure case discussed above, and the corresponding tracks measured on BCRs and CCRs, used to reach the conclusions of this study is tabulated in Table~S1. For each exposure time $t$, the difference between the mean BCR track counts and the mean CCR counts representing, the slow-neutron-specific signal ($D$), defined as  

\begin{equation}
	D = B - C,
	\tag{1}
	\label{eq:1}
\end{equation}

where $B$ and $C$ are the mean track counts for the BCR and CCR detectors, respectively. The associated uncertainty $\sigma_D$ was obtained by propagation of the independent standard errors of the mean, $\sigma_B$ and $\sigma_C$, corresponding to the BCR and CCR track counts, respectively.

\begin{equation}
	\sigma_{D} = \sqrt{\sigma_B^{2} + \sigma_C^{2}} .
	\tag{2}
	\label{eq:2}
\end{equation}
\\
Clearly, for all exposure times, Table 1 shows that the BCR and CCR detectors recorded significant number of tracks, and the difference between them seems to be increasing linearly with increase in exposure time. The measured data points $(t,\, D \pm \sigma_D)$, of Table~I,therefore provides the required information for determining the BCR detector neutron detection rate discussed below.

To evaluate the slow-neutron detection rate of BCR, assumed to be a constant $R$, is determined using a linear model of the form

\begin{equation}
	D = R t ,
	\tag{3}
	\label{eq:3}
\end{equation}

where $D$ denotes BCR--CCR track difference, $t$ is the exposure time and and R is the constant slope to be determined from a linear fit. The validity of this constant-rate assumption and the consistency of the model with the experimentally measured data are quantitatively assessed using a chi-square ($\chi^2$) goodness-of-fit analysis, providing a numerical measure of the agreement between the linear model and the observed data.
\\

\onecolumngrid

\clearpage

\begin{table}[h]
	\caption{Track-count data obtained on boron-coated CR-39 (BCR) and control CR-39 (CCR) detectors for different irradiation times $t$.  For each exposure time $t$, the BCR/CCR counts obtained on its four non--overlapping regions is mentioned in a sequence: LT, RT, RB, and LB. Each data point in the column labeled $t$ shares a common standard error of $\sigma_t = 0.5$~min.}
	\label{tab:S1}
	\begin{ruledtabular}
		\setlength{\tabcolsep}{6pt}
		\renewcommand{\arraystretch}{1.25}
		
		\begin{tabular}{c c c c c c}
			\begin{tabular}[c]{@{}c@{}}
				$t$\\
				(min)
			\end{tabular} &
			BCR counts &
			\begin{tabular}[c]{@{}c@{}}
				Mean BCR counts\\
				$(B \pm \sigma_B)$
			\end{tabular} &
			CCR counts &
			\begin{tabular}[c]{@{}c@{}}
				Mean CCR counts\\
				$(C \pm \sigma_C)$
			\end{tabular} &
			\begin{tabular}[c]{@{}c@{}}
				$D \pm \sigma_D$\\
				$((B \pm \sigma_B)-(C \pm \sigma_C))$
			\end{tabular} \\
			\hline
			10 & 140, 144, 140, 141 & $141.25 \pm 0.95$ & 89, 87, 80, 62 & $79.50 \pm 6.14$ & $61.75 \pm 6.21$ \\
			20 & 250, 283, 276, 250 & $264.75 \pm 8.51$ & 140, 130, 149, 141 & $140.00 \pm 3.98$ & $124.75 \pm 9.39$ \\
			30 & 418, 366, 335, 376 & $373.75 \pm 17.45$ & 211, 198, 189, 183 & $195.25 \pm 6.01$ & $178.50 \pm 18.46$ \\
			40 & 507, 504, 505, 533 & $512.25 \pm 6.57$ & 288, 270, 290, 283 & $282.75 \pm 4.12$ & $229.50 \pm 7.76$ \\
		\end{tabular}
		
	\end{ruledtabular}
\end{table}
\twocolumngrid
\clearpage
We first note that the measurement of exposure time $t$ involves an uncertainty $\sigma_t = 0.5$~min, and this uncertainty propagates into the slow-neutron signal $D$ as

\begin{equation}
	\delta D = R\,\sigma t .
	\tag{4}
	\label{eq:4}
\end{equation}

Accordingly, the effective variance associated with each data point D, as per standard error propagation, is given by

\begin{equation}
	\sigma_{\mathrm{eff},i}^{2} = \sigma_{D_i}^{2} + R^{2}\sigma_t^{2} ,
	\tag{5}
	\label{eq:5}
\end{equation}

where $i$ is a running index (i=1-4) corresponding to the four measured track datasets obtained at the four different exposure times listed in Table~I.

The best-fit value of $R$ is obtained by minimizing the full chi-square $\chi^{2}(R)$, defined as

\begin{equation}
	\chi^{2}(R) =
	\sum_{i}
	\frac{(D_i - R t_i)^{2}}
	{\sigma_{D_i}^{2} + R^{2}\sigma_t^{2}} .
	\tag{6}
	\label{eq:6}
\end{equation}

Here, $D_i$ are the measured quantities (BCR--CCR differences), $t_i$ are the measured exposure times, $\sigma_{D_i}$ arise from spatial SEM statistics, and $\sigma_t$ is the known timing uncertainty.


Neglecting the time uncertainty ($\sigma_t \rightarrow 0$), the chi-square in equation S6 reduces to

\begin{equation}
	\chi_{0}^{2}(R) =
	\sum_{i}
	\frac{(D_i - R t_i)^{2}}
	{\sigma_{D_i}^{2}} .
	\tag{7}
	\label{eq:7}
\end{equation}

(Here, $\chi_0^2$ denotes the maximum-likelihood form obtained when the uncertainty in the exposure time is neglected, assuming Gaussian statistics for $D_i$)

For minimum $\chi_{0}^{2}$,
\begin{equation}
	\frac{d\chi_{0}^{2}}{dR}
	=
	\sum_{i}
	\frac{2(R t_i - D_i)t_i}
	{\sigma_{D_i}^{2}}
	= 0 .
	\tag{8}
	\label{eq:8}
\end{equation}

Rearranging,
\begin{equation}
	R \sum_{i} \frac{t_i^{2}}{\sigma_{D_i}^{2}}
	=
	\sum_{i} \frac{t_i D_i}{\sigma_{D_i}^{2}} .
	\tag{9}
	\label{eq:9}
\end{equation}

Solving for $R$ gives the unique maximum-likelihood estimator
\begin{equation}
	R
	=
	\frac{\sum_{i} t_i D_i / \sigma_{D_i}^{2}}
	{\sum_{i} t_i^{2} / \sigma_{D_i}^{2}}
	\approx 5.85~\text{tracks/min},
	\tag{10}
	\label{eq:10}
\end{equation}
using data from Table~S1.
The obtained slope, $R = 5.85~\text{tracks}\,\text{min}^{-1}$, is used as the initial value for the iterative fitting procedure and is hereafter denoted as $R_0$.

The effective variance using the initial slope is therefore
\begin{equation}
	(R_{0}\sigma_t)^{2} = (5.85 \times 0.5)^{2} = 8.56 ,
	\tag{11}
	\label{eq:11}
\end{equation}
and hence
\begin{equation}
	\sigma_{\mathrm{eff}}^{2} = \sigma_{D}^{2} + 8.56 .
	\tag{12}
	\label{eq:12}
\end{equation}

As per equation S12, the evaluated quantities to be used for the $\chi^{2}$ minimization to obtain the slope of next iteration ($R_{1}$) is summarized in Table~II.

\begin{table}[H]
	\caption{Values used in the $\chi^{2}$ minimization procedure.}
	\label{tab:S2}
	\begin{ruledtabular}
		\setlength{\tabcolsep}{6pt}
		\renewcommand{\arraystretch}{1.2}
		\begin{tabular}{c c c c }
			$t_i$ (min) & $D_i$ & $\sigma_{D_i}^{2}$ & $\sigma_{\mathrm{eff},i}^{2}$  \\
			\hline
			10 & 61.75 & 38.56 & 47.12 \\
			20 & 124.75 & 88.20 & 96.76  \\
			30 & 178.50 & 341.0 & 349.56  \\
			40 & 229.50 & 60.20 & 68.76  \\
		\end{tabular}
	\end{ruledtabular}
\end{table}

Using the full $\chi^{2}$ weights, the updated slope $R_{1}$, using the data in table~II is obtained as
\begin{equation}
	R_{1}
	=
	\frac{\sum_{i} t_i D_i / \sigma_{\mathrm{eff},i}^{2}}
	{\sum_{i} t_i^{2} / \sigma_{\mathrm{eff},i}^{2}} .
	\tag{13}
	\label{eq:13}
\end{equation}

A second iteration changes slope $R$ by less than $0.01$, confirming convergence. Explicitly,
\begin{equation}
	\lvert R_{1} - R_{0} \rvert = 0.01 \ll \sigma_{R} = 0.18~\&~ R^{2}\sigma_{t}^{2} \ll \sigma_{D_i}^{2}.
	\tag{14}
	\label{eq:14}
\end{equation}

(Calculation details of $\sigma_{R}$ is provided in Sec. II of Supplemental material). The fitted slow-neutron detection rate is therefore
\begin{equation}
	R = 5.84 \pm 0.18~\text{tracks}\,\text{min}^{-1} .
	\tag{15}
	\label{eq:15}
\end{equation}

The calculated values of the effective variance (equation 5) with the slope value of $R = 5.84 \pm0.18~\text{tracks}\,\text{min}^{-1}$, used in the $\chi^{2}$ evaluation where
\begin{equation}
	R^{2}\sigma_{t}^{2}
	=
	(5.84 \times 0.5)^{2}
	=
	8.56 .
	\tag{16}
	\label{eq:16}
\end{equation}
turns out to be identical to those listed in Table~II, validating that the fit slope has converged.

\subsection{Model predictions and the statistical significance of the fitted rate R}

Using the linear model,
\begin{equation}
	D_{\mathrm{model},i} = R t_i ,
	\tag{17}
	\label{eq:17}
\end{equation}
the individual $\chi_i^{2}$ contributions are
\begin{equation}
	\chi_i^{2}
	=
	\frac{(D_i - R t_i)^{2}}
	{\sigma_{\mathrm{eff},i}^{2}} .
	\tag{18}
	\label{eq:18}
\end{equation}

The total chi-square is
\begin{equation}
	\chi^{2}_{\min}
	=
	0.238 + 0.653 + 0.031 + 0.244
	=
	1.17 .
	\tag{19}
	\label{eq:19}
\end{equation}

The number of data points is $N = 4$, and the number of fitted parameters is $p = 1$.  
Thus, the degrees of freedom are
\begin{equation}
	\nu = N - p = 3 .
	\tag{20}
	\label{eq:20}
\end{equation}

The reduced chi-square is
\begin{equation}
	\chi_{\mathrm{red}}^{2}
	=
	\frac{\chi^{2}_{\min}}{\nu}
	=
	\frac{1.17}{3}
	=
	0.39 .
	\tag{21}
	\label{eq:21}
\end{equation}

The condition $\chi_{\mathrm{red}}^{2} \ll 1$ indicates excellent consistency with the linear model, confirms that the assigned uncertainties (Standard error of means and timing uncertainty) are not underestimated, and shows no evidence for curvature (No time-dependent enhancement or suppression of detection rate), saturation (No time-dependent enhancement or suppression of detection rate), or drift in the time dependence (detection rate does not ages with time.). The detector response is stable over the full exposure interval.

Furthermore, the statistical significance of the fitted rate as per equation~15. is
\begin{equation}
	Z = \frac{R}{\sigma_R}
	=
	\frac{5.84}{0.18}
	\approx 32 .
	\tag{22}
	\label{eq:22}
\end{equation}

The BCR--CCR signal therefore increases linearly with exposure time. Using SEM-based uncertainties, full propagation of a 0.5-min timing error, and iterative $\chi^{2}$ minimization, the fitted neutron-equivalent detection rate is
\begin{equation}
	R = (5.84 \pm 0.18)\,\text{tracks}\,\text{min}^{-1}\,(4.8~\text{mm}^{2})^{-1},
	\tag{23}
	\label{eq:23}
\end{equation}

differing from zero at the $\sim32\sigma$ level which is also a result of this study .

\subsection{Slow neutron detection efficiency of BCR}
The slow neutron detection efficiency of the boron-coated CR-39 (BCR), $\varepsilon$, is calculated as the ratio of the particle tracks recorded on the BCR per minute of exposure,
$R = (5.84 \pm 0.18)\,\text{tracks}\,\text{min}^{-1}\,(4.8~\text{mm}^{2})^{-1}$,
to the total number of thermal neutrons incident on the detector during the same time interval,
$N = (3658 \pm 183)\,\text{neutrons}\,\text{min}^{-1}\,(4.8~\text{mm}^{2})^{-1}$,
assuming a $5\%$ uncertainty in the source flux~\cite{PURTY2017144}.
Accordingly, the detection efficiency is obtained as
$\varepsilon \simeq (1.60 \pm 0.09)\times 10^{-3}$.

\subsection{Experimental significance}
\subsubsection {BCR-CCR as an suitable detector system for slow neutron measurements}
The first outcome of this study arises from the observed stability of the neutron detection rate R, which indicates a constant detection efficiency and thereby confirms the suitability of the fabricated BCR–CCR detector system for quantitative slow-neutron flux measurements

\subsubsection{Measurement of the slow-to-fast neutron flux ratio in mixed neutron fields}
An important outcome of this study, relevant to thermal neutron sources that are commonly accompanied by fast-neutron leakage, is that the associated fast-neutron contribution can be quantitatively characterized through the tracks recorded on the CCR component of the BCR–CCR detector system. This provides direct information on fast-neutron leakage from the thermal neutron source and enables an estimate of the thermal-to-fast neutron flux ratio, provided that the fast-neutron detection efficiency of CR-39 is known.

Additional advantages of the BCR–CCR system become evident when compared with earlier approaches used to separate thermal neutron fluxes in mixed fast–slow neutron fields. Previous studies have commonly employed the cadmium-difference method, which relies on comparative measurements performed with and without a cadmium cover placed over a boron-coated CR-39 detector. This technique exploits the large neutron absorption cross section of cadmium for neutron energies below approximately $0.5$ eV \cite{smilgys2013boron}.

 However, the method is inherently affected by distortion of the slow-neutron energy spectrum near the cadmium cutoff and by perturbations of the local neutron field arising from additional scattering and absorption introduced by the Cd layer. These effects can lead to systematic uncertainties and inaccuracies in the differential estimation of slow-neutron fluence.

An alternative study employed a boron integrated CR-39, with slow neutron signals being recorded on boron coated side and fast neutron contributions estimated from recoil-proton tracks on the reverse side of the detector surface. Although effective, this technique considers symmetric detector response and is sensitive to converter thickness,scattering and absorption of fast neutrons effects~\cite{MAMELI20083656}.

In contrast, the differential BCR–CCR method discussed in this study eliminates the need for external absorbers and dual-surface analysis. Simultaneous exposure and identical processing of boron-coated and bare CR-39 detectors enable intrinsic compensation of fast-neutron and background contributions, thereby providing a more robust and spectrally faithful determination of the slow neutron flux in mixed-fast neutron environments.
\subsubsection{Slow neutron measurement in an unknown field}
The most important outcome of this study is the establishment of a robust measurement methodology for identifying potential slow-neutron fluxes in radiation fields of unknown composition. For example, experimental investigations of high-voltage air discharges have reported particle tracks on boron-coated CR-39 detectors that were attributed to slow-neutron emission. In such systems, where the underlying emission mechanism is not well understood, the possible generation of charged particles and/or fast neutrons must be carefully considered, as these components can also produce tracks in boron-integrated CR-39 detectors and thereby lead to overestimation of slow-neutron fluence. We therefore emphasize that simultaneous deployment of an uncoated CR-39 control detector alongside the boron-based thermal-neutron sensor is essential, as it enables direct measurement of non-thermal background contributions and allows a more accurate and physically consistent determination of slow-neutron flux. A detailed demonstration of this methodology in liquid-medium experiments has been reported in ~\cite{kumar2026cr39trackdetectorsignatures}, where the observed differential track signal was statistically significant while the CCR response remained near background levels, indicating the dominant presence of slow neutrons and thereby demonstrating the versatility of the approach. The technique is likewise well suited for measuring the slow-neutron component of radiation fields in plasma systems, particularly in configurations where CR-39 detectors have already been employed for diagnostics of fusion byproducts ~\cite{lahmann2020cr,seguin2003spectrometry}.

\section{Conclusions}
  This study demonstrates that boron-coated CR-39 detectors exhibit a linear and statistically robust response to thermal neutron exposure over the investigated range. Poisson-based uncertainty analysis and statistical consistency tests confirm that the detector response is compatible with a constant neutron flux and stable detection efficiency, thereby validating the use of BCR-CCR detectors for quantitative thermal-neutron measurements under controlled irradiation conditions.
   
  Accurate thermal-neutron measurements in complex media including liquids, electrochemical cells, vacuum systems, plasma discharges, and space-constrained geometries remain experimentally challenging, as conventional neutron detectors are often difficult or impractical to deploy. In such environments, mixed radiation fields and low neutron fluences introduce substantial ambiguity in track-based measurements, limiting the reliability of standalone CR-39 detectors.
  
  The principal novelty of the present work lies in the implementation of a paired BCR–CCR differential detection methodology. By deploying a boron-coated CR-39 detector together with an uncoated CR-39 control at the same detection site, background contributions arising from charged particles and fast neutrons are quantified directly through the CCR response. Subtraction of this background from the BCR signal yields a conservative and internally self consistent estimate of slow-neutron fluence. This differential approach enables unambiguous identification of thermal neutron induced tracks at levels comparable to intrinsic detector background, thereby overcoming a key limitation of conventional CR-39 based neutron measurements in mixed radiation environments.

\begin{acknowledgments}
The authors acknowledges the contributions of Sandir Soy for his diligent efforts in calibration data collection and lab assistance. 
\end{acknowledgments}

\bibliography{apssamps}


\clearpage
\onecolumngrid 
\clearpage
\begin{center}
	\textbf{\large Supplemental Material to “Experimental Determination of Slow-Neutron Detection Efficiency and Background Discrimination in Mixed Radiation Fields Using Differential CR-39 Track Detectors”
	\\
	\vspace{0.2em}
	by Ankit Kumar, Tushar Verma, Pankaj Jain, Raj Ganesh Pala and K.~P.~Rajeev}
\end{center}
\vspace{-0.5cm}
\setcounter{equation}{0}
\setcounter{figure}{0}
\setcounter{table}{0}
\setcounter{section}{0}
\renewcommand{\theequation}{S\arabic{equation}}
\renewcommand{\thefigure}{S\arabic{figure}}
\renewcommand{\thetable}{S\arabic{table}}

\vspace{0.8cm}
\maketitle

The raw data involved in this study is provided in Section~I. The uncertainty calculation on R value is provided in Section~II. Figures, equations, and tables appearing in this Supplemental Material are labeled S1, S2, \dots, to distinguish them from those in the main article.

\section{Raw Data for BCR and CCR exposed to reference thermal neutron source} 
\begin{center}
	\includegraphics[width=\textwidth]{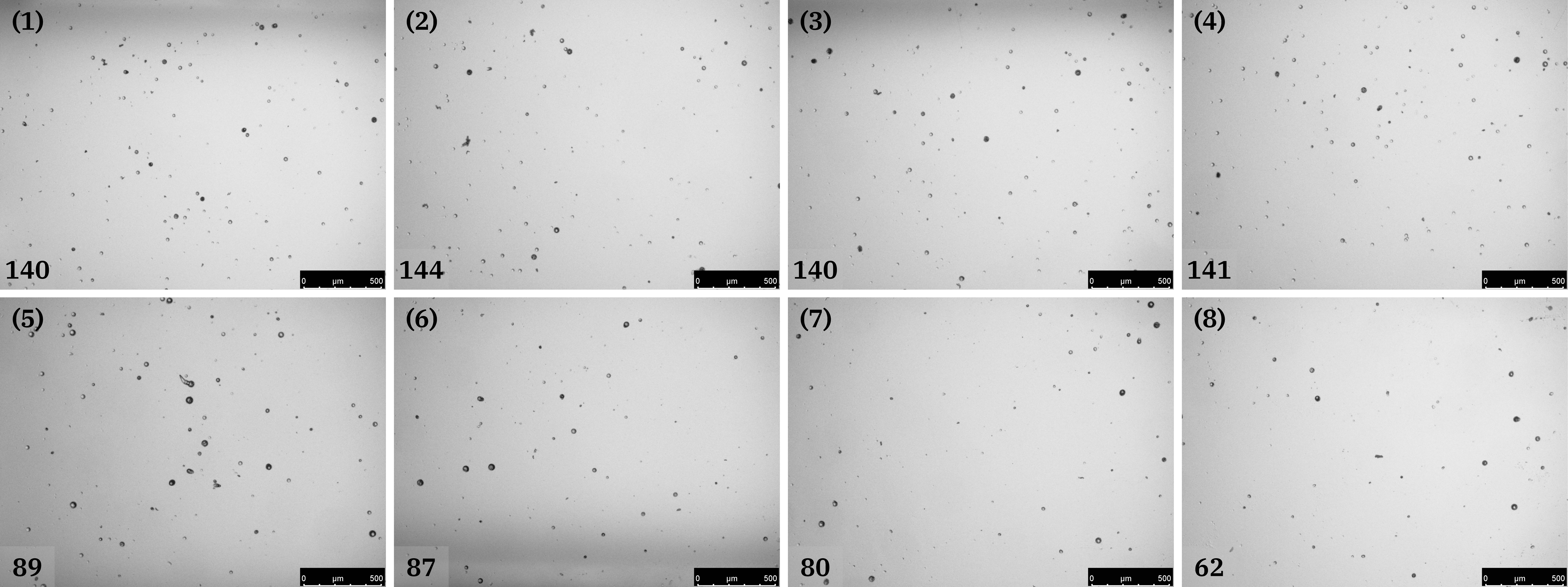}\\[0.2em]
	\includegraphics[width=\textwidth]{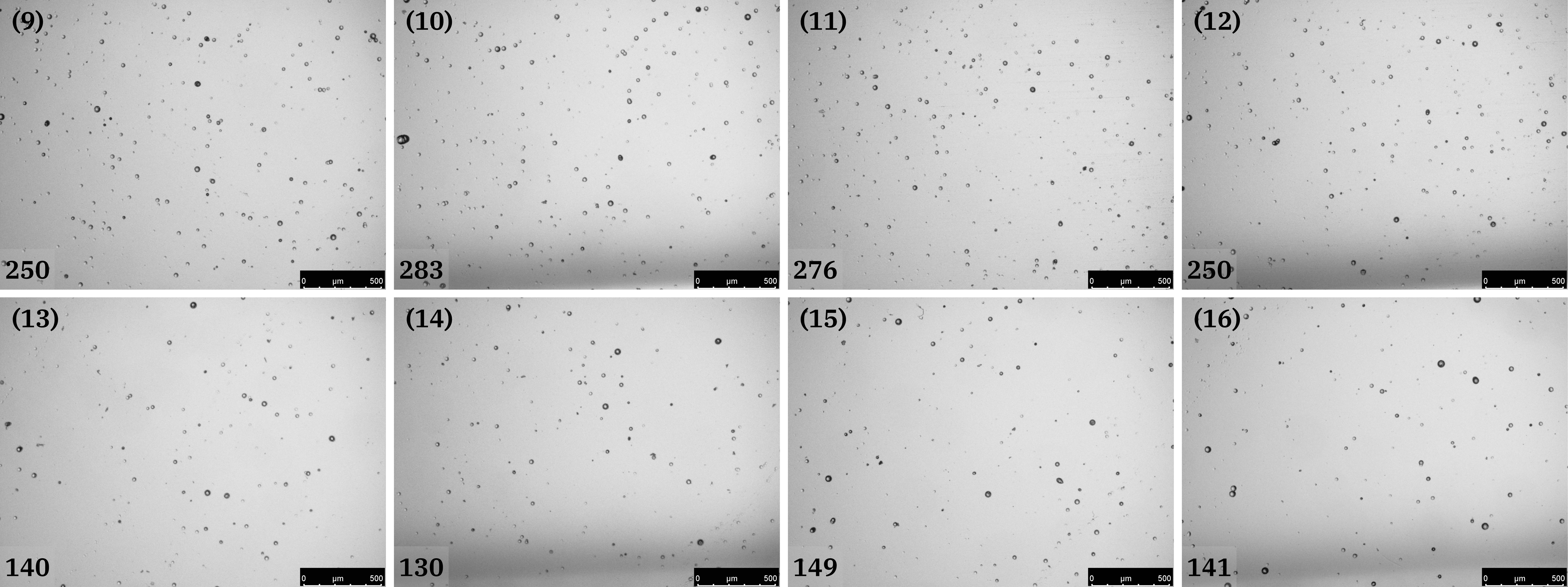}\\[0.2em]
	\includegraphics[width=\textwidth]{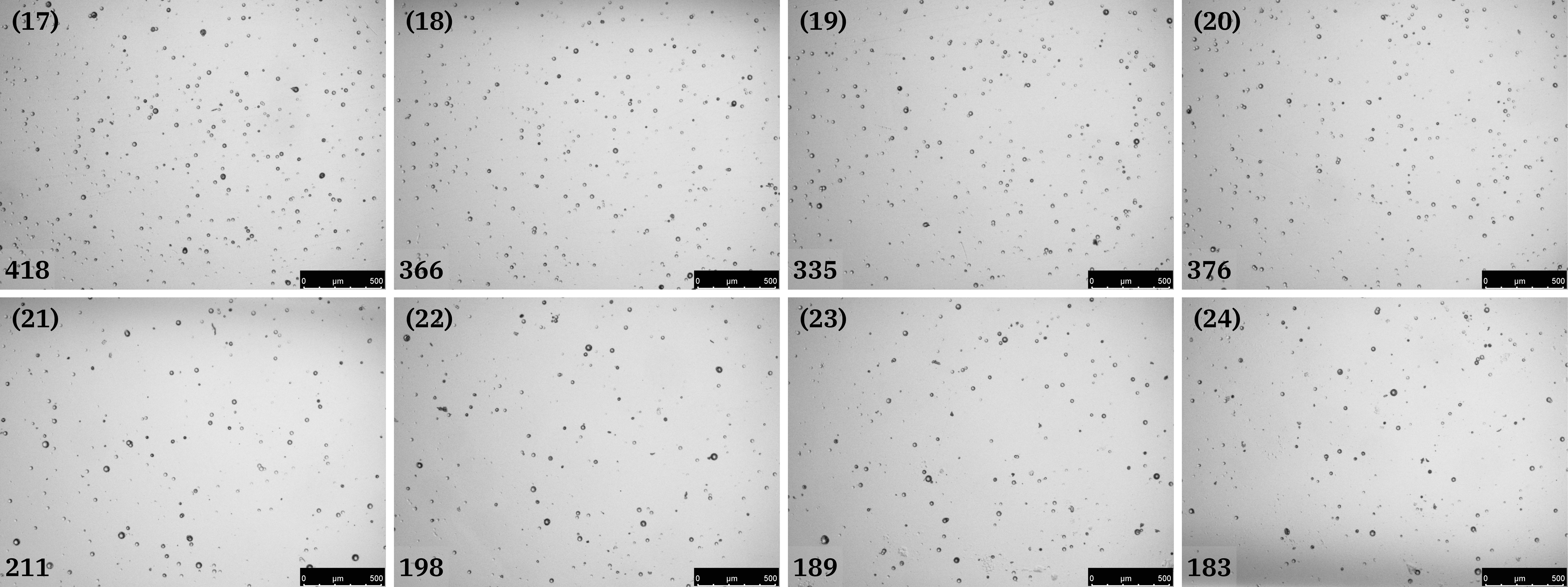}\\[0.2em]
	\includegraphics[width=\textwidth]{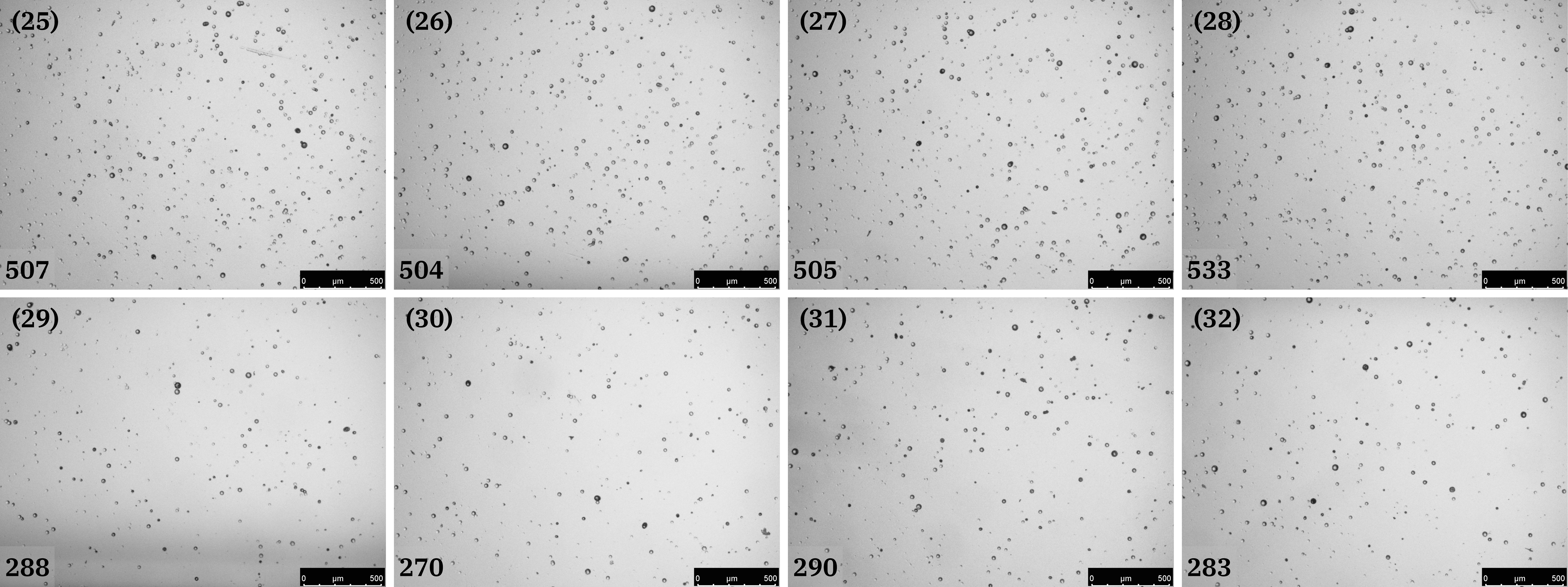}\\[0.2em]
	
	\captionsetup{type=figure, width=\linewidth, justification=justified, format=plain}
	\captionof{figure}{\justifying Optical micrographs of boron-coated CR-39 (BCR) and uncoated control CR-39 (CCR) detectors exposed to a reference thermal neutron source for different durations are shown. Each row of images corresponds to a single BCR/CCR detector pair. From left to right, the panels represent the left-top (LT), right-top (RT), right-bottom (RB), and left-bottom (LB) regions of the detector surface. The image labeling is as follows: 10-min exposure---BCR (images 1--4) and CCR (images 5--8); 20-min exposure---BCR (images 9--12) and CCR (images 13--16); 30-min exposure---BCR (images 17--20) and CCR (images 21--24); and 40-min exposure---BCR (images 25--28) and CCR (images 29--32).}
	\label{fig:S1}
\end{center}

\twocolumngrid
\clearpage
\section{Calculation of Uncertainity on RATE CONSTANT/SLOPE VALUE R}

For the uncertainty on $R$, represented by $\sigma_{R}$, the $\chi^{2}(R)$ function is expanded as a Taylor series about its minimum,
\begin{equation}
	\chi^{2}(R)
	\approx
	\chi^{2}_{\min}
	+
	\frac{1}{2}
	\left.
	\frac{d^{2}\chi^{2}}{dR^{2}}
	\right|_{R = R_{\min}}
	(R - R_{\min})^{2} .
	\tag{S1}
	\label{eq:S1}
\end{equation}

By definition of Gaussian confidence intervals, the $1\sigma$ uncertainty on $R$ corresponds to the value of $R$ for which
\begin{equation}
	\chi^{2}(R) = \chi^{2}_{\min} + 1 .
	\tag{S2}
	\label{eq:S2}
\end{equation}

Hence,
\begin{equation}
	\chi^{2}(R) - \chi^{2}_{\min}
	\approx
	\frac{1}{2}
	\left.
	\frac{d^{2}\chi^{2}}{dR^{2}}
	\right|_{R_{\min}}
	(R - R_{\min})^{2} .
	\tag{S3}
	\label{eq:S3}
\end{equation}

Setting $\chi^{2}(R) - \chi^{2}_{\min} = 1$ and $(R - R_{\min})^{2} = \sigma_{R}^{2}$ gives
\begin{equation}
	1 =
	\frac{1}{2}
	\left.
	\frac{d^{2}\chi^{2}}{dR^{2}}
	\right|_{R_{\min}}
	\sigma_{R}^{2} .
	\tag{S4}
	\label{eq:S4}
\end{equation}

Solving for $\sigma_{R}$,
\begin{equation}
	\sigma_{R}
	=
	\sqrt{
		\frac{2}
		{\left.
			\frac{d^{2}\chi^{2}}{dR^{2}}
			\right|_{R_{\min}}}
	} .
	\tag{S5}
	\label{eq:S5}
\end{equation}

For this study,
\begin{equation}
	\chi^{2}(R)
	=
	\sum_{i} w_i (D_i - R t_i)^{2},
	\qquad
	w_i = \frac{1}{\sigma_{\mathrm{eff},i}^{2}} ,
	\tag{S6}
	\label{eq:S6}
\end{equation}
and the second derivative is
\begin{equation}
	\frac{d^{2}\chi^{2}}{dR^{2}}
	=
	2 \sum_{i} w_i t_i^{2} .
	\tag{S7}
	\label{eq:S7}
\end{equation}

Thus,
\begin{equation}
	\sigma_{R}
	=
	\frac{1}{\sqrt{\sum_{i} t_i^{2} w_i}}
	=
	\frac{1}{\sqrt{32.09}}
	=
	0.18 .
	\tag{S8}
	\label{eq:S8}
\end{equation}

The uncertainty on $R$ is dominated by the spread in the 40-min data point, which carries the largest statistical weight. The inclusion of the time uncertainty ($\sigma_t = 0.5$~min) increases $\sigma_{R}$ modestly but does not affect convergence. Because $\chi_{\mathrm{red}}^{2} < 1$, the quoted uncertainty is conservative.




\end{document}